\documentclass[twocolumn,showpacs,prc,aps,floatfix,reprint]{revtex4-1}
\usepackage{amssymb}
\usepackage{braket}
\usepackage{graphicx}
\usepackage{dcolumn}
\usepackage{amsmath}
\usepackage{amsfonts}
\usepackage{color}
\usepackage{bm}
\begin{document}

\title{Validity of the distorted-wave impulse-approximation description of ${}^{40}$Ca$(e,e'p){}^{39}$K data using only ingredients from a nonlocal dispersive optical model}
\author{M. C. Atkinson$^1$, H.P. Blok$^{2,3}$, L. Lapik\'{a}s$^{2}$, R. J. Charity$^4$, and W. H. Dickhoff$^1$} 
\affiliation{${}^1$Department of Physics,
Washington University, St. Louis, Missouri 63130, USA}
\affiliation{${}^2$National Institute for Subatomic Physics (Nikhef), 1009 DB Amsterdam, The Netherlands}
\affiliation{${}^3$Department of Physics and Astronomy, VU University, 1081 HV Amsterdam, The Netherlands}
\affiliation{${}^4$Department of Chemistry,
Washington University, St. Louis, Missouri 63130, USA}

\date{\today}

\begin{abstract}
The nonlocal implementation of the dispersive optical model (DOM) provides all the ingredients for distorted-wave impulse-approximation (DWIA) calculations of the $(e,e'p)$ reaction. It provides both the overlap function, including its normalization, and the outgoing proton distorted wave.
This framework is applied to describe the knockout of a proton from the $0\textrm{d}\frac{3}{2}$ and $1\textrm{s}\frac{1}{2}$ orbitals in ${}^{40}$Ca with fixed normalizations of 0.71 and 0.60, respectively. Data were obtained in parallel kinematics for three outgoing proton energies: 70, 100, and 135 MeV.
Agreement with the data is as good as, or better than, previous descriptions employing local optical potentials and overlap functions from Woods-Saxon potentials - both with standard nonlocality corrections - whose normalization (spectroscopic factor) and radius were fitted to the data.
The present analysis suggests that slightly larger spectroscopic factors are obtained when nonlocal optical potentials are employed than those generated with local potentials.
The results further suggest that the chosen kinematical window around 100 MeV proton energy provides the best and cleanest method to employ the DWIA for the analysis of this reaction. 
The conclusion that substantial ground-state correlations cannot be ignored when describing a closed-shell atomic nucleus is therefore confirmed in detail.
To reach these conclusions, it is essential to have a complete description of the nucleon single-particle propagator that accounts for all elastic nucleon-scattering observables in a wide energy domain up to 200 MeV. The current nonlocal implementation of the DOM fulfills this requirement.\end{abstract}

\maketitle

\section{Introduction}

The shell model, in which the nucleons fill certain orbitals, is well suited to describe the structure of a nucleus. The best place to test this description is in or around (double) closed-shell nuclei. In the simplest picture, in which residual interactions are neglected, all orbitals are 100\% filled up to the Fermi level according to the Pauli principle, and those above it are empty. 
However, due to residual interactions there is depletion of orbitals below the Fermi energy, and filling of those above it.
The precise amount of this depletion/filling is still a topic under investigation.
   The best tool to study this experimentally is the $(e,e'p)$ reaction~\cite{Kramer:1989,Denherder:1988,Peter90,Lex90,Ingo91,Lapikas93,Pandharipande97}. 
   At sufficiently high electron energy and momentum transfer, the proton can be knocked out with enough energy such that a description within the distorted-wave impulse approximation (DWIA)  can be expected to be applicable, so that depletion (and also filling) of orbits can be studied.
   
   The canonical analysis of this reaction, practiced by the Nikhef group~\cite{Denherder:1988,Kramer:1989},
   employs a standard global optical potential for the distorted wave and calculates the bound-state wave function (overlap function) of the proton in a Woods-Saxon potential well, which is adjusted  to describe the momentum dependence of the measured cross sections.
   A scaling factor of about 0.6-0.7 (relative to a completely filled orbital) is then required to describe the overall magnitude~\cite{Lapikas93}.
   This scaling factor corresponds to the normalization of the overlap function between
   the target ground state and low-lying single-hole states, usually referred to as the (reduced) spectroscopic factor.
   Often this spectroscopic factor is reported multiplied with a factor of $2j+1$ corresponding to the complete filling of a shell with angular momentum $j$. 
Furthermore, the data show that additional removal strength with essentially the same overlap function is located at nearby energies, providing clear evidence of the fragmentation of the single-particle strength~\cite{Kramer:1989,Kramer90}.
   
%
   
 The theoretical interpretation of these experimental results, reviewed in Refs.~\cite{Pandharipande97,Dickhoff04}, has mainly been concerned with the explanation of this reduction in the spectroscopic strength to 60-70\% of the independent-particle shell model (IPSM) value. 
Whereas the main reduction of the strength appears to be due to the coupling to low-lying surface vibrations and higher-lying giant resonances associated with long-range correlations (LRC), it has been well documented that additional short-range and tensor correlations (SRC) can be responsible for a 10-15\% depletion of the IPSM value~\cite{Dickhoff04}.

The interpretation of spectroscopic factors has been questioned in the literature~\cite{Zhanov10,Furnstahl10,Jennings11} as well as the possibility of measuring momentum distributions or occupation probabilities~\cite{Furnstahl02}.
In order to address this issue, it is useful to rephrase the interpretation of the $(e,e'p)$ cross section as a question whether the DWIA is a valid reaction model for this process.
It is also important to describe the data with a consistent set of ingredients.
For example, in the standard Nikhef analysis the potential used to describe the distorted proton wave is unrelated to the one that generates the overlap function and its normalization is a scaling parameter to fit the data.

Recent developments of the Dispersive Optical Model (DOM) make it possible to provide all the necessary ingredients of the DWIA for this reaction.
The DOM was developed by Mahaux and Sartor~\cite{Mahaux91} to provide the link between the potential used to describe elastic nucleon scattering data and the one that provides the levels of the IPSM through the use of a subtracted dispersion relation, which links the real and imaginary parts of the nucleon self-energy~\cite{Exposed!}.
Recent implementations of the DOM have introduced fully nonlocal potentials~\cite{Mahzoon:2014,Dickhoff:2017} to allow additional data to be included in the description, like particle number and the nuclear charge density.
It is thus possible to provide all the ingredients of the DWIA from the DOM description of all available elastic nucleon scattering data as well as separation energies, particle number, and the nuclear charge density for ${}^{40}$Ca in our case.
Indeed, the distorted outgoing proton wave and the overlap function with its implied normalization are all provided by the DOM to allow a consistent description of the
$^{40}$Ca$(e,e'p)$$^{39}$K cross section for the three available energies of 70, 100, and 135 MeV of the outgoing proton. 
   The states analyzed for this reaction are the first two states of $^{39}$K, corresponding to the $0\textrm{d}\frac{3}{2}$ and $1\textrm{s}\frac{1}{2}$ valence hole states in $^{40}$Ca in the IPSM. 
   The three different proton energies were
   chosen to test the validity of the DWIA used to calculate the theoretical cross sections for this range of energies,  which involves the folding of the ejected proton's bound-state wave function (overlap function with the appropriate normalization) with its outgoing distorted wave to calculate the cross section~\cite{Giusti:1988,Boffi:1980}. 
   
   In the past, the spectroscopic factor was found by scaling the calculated cross sections to match the data. 
   In the present analysis, the DOM also provides the spectroscopic factor allowing a consistent description of the cross section and thereby an assessment of the accuracy of the DWIA description.
   In addition, it is possible to check the consistency between the data that determine the DOM self-energy and the $(e,e'p)$ cross sections.
   
   In Sec.~\ref{sec:theory} we review the theoretical ingredients of the analysis.
   It includes in Sec.~\ref{sec:prop} the relevant material related to the Green's function method that provides the framework of the discussion.
   In Sec.~\ref{sec:DOM} essential ingredients of the DOM are presented, while the DWIA of the $(e,e'p)$ cross section is described in Sec.~\ref{sec:DWIA}.
   The experiment is described in Sec.~\ref{sec:EXP} and the results presented in Sec.~\ref{sec:results}.
   Conclusions and outlook are discussed in Sec.~\ref{sec:CON}.

      \section{Theory}
\label{sec:theory}
   The DOM generates all of the ingredients necessary to calculate the $(e,e'p)$ cross section when the DWIA is adopted. It provides a representation of the nucleon self-energy that is constrained by a large number of observables related to adding or removing a particle from the ground state, ${}^{40}$Ca in this case.
It employs the framework of the Green's function method~\cite{Exposed!} to simultaneously describe all available elastic nucleon scattering cross sections as well as neutron and proton particle number, removal energies of discrete valence orbits below the Fermi energy, and the nuclear charge density. In addition, it provides relevant quantities for the analysis of reactions, including overlap functions with their normalization and distorted waves for nucleons at positive energy.
This section is broken up in subsections that provide brief introductions to all the concepts used in the present analysis.

   \subsection{Single-particle propagator}
\label{sec:prop}
   The single-particle propagator describes the probability amplitude for adding a particle in state $\alpha$ at one time to the ground state of a system and propagating on top of that state until a later time at which it is removed
  in state $\beta$~\cite{Exposed!}.  In addition to the conserved orbital and total angular momentum ($\ell$ and $j$, respectively), the labels $\alpha$ and $\beta$ in Eq.~(\ref{eq:green}) refer to a suitably chosen single-particle basis. 
 In this work the Lagrange basis~\cite{Baye:2010} was employed.
It is convenient to work with the Fourier-transformed propagator in the energy domain, 
   \begin{eqnarray}
  \!\!\! \!   G_{\ell j}(\alpha,\beta;E) \!\! = \!\! \bra{\Psi_0^A}a_{\alpha \ell j} \frac{1}{E-(\hat{H}-E_0^A)+i\eta} a_{\beta \ell j}^\dagger\ket{\Psi_0^A} \nonumber \\*
   \!\!\!\!\!  + \! \bra{\Psi_0^A}a_{\beta \ell j}^\dagger\frac{1}{E-(E_0^A-\hat{H})-i\eta} a_{\alpha \ell j}\ket{\Psi_0^A},
      \label{eq:green}
   \end{eqnarray}
   with $E^A_0$ representing the energy of the nondegenerate ground state $\ket{\Psi^A_0}$.
   Many interactions can occur between the addition and removal of the particle (or \textit{vice versa}), all of which need to be considered to calculate the propagator. 
No assumptions about the detailed form of the Hamiltonian $\hat{H}$ need to be made for the present discussion, but it will be assumed that a meaningful Hamiltonian exists that contains two-body and three-body contributions.
 Application of perturbation theory then leads to the Dyson equation~\cite{Exposed!} given by
   \begin{eqnarray}
      G_{\ell j}(\alpha,\beta;E) = G_{\ell}^{(0)}(\alpha,\beta;E) \qquad\qquad\qquad\qquad \qquad\qquad \nonumber 
      \\* 
      + \sum_{\gamma,\delta}G_{\ell}^{(0)}(\alpha,\gamma;E)\Sigma_{\ell j}^*(\gamma,\delta;E)G_{\ell j}(\delta,\beta;E) , \qquad 
      \label{eq:dyson}
   \end{eqnarray}
   where $G^{(0)}_{\ell}(\alpha,\beta;E)$ corresponds to the free propagator (which only includes a kinetic contribution)
   and $\Sigma_{\ell j}^*(\gamma,\delta;E)$ is the irreducible self-energy~\cite{Exposed!}. 

   The hole spectral density for energies below $\varepsilon_F$ is obtained from 
   \begin{equation}
      S^h_{\ell j}(\alpha,\beta;E) = \frac{1}{\pi}\textrm{Im}\ G_{\ell j}(\alpha,\beta;E) .
      \label{eq:spec}
   \end{equation}
   The diagonal element of Eq.~(\ref{eq:spec}) is known as the (hole) spectral function identifying the probability density for the removal of a single-particle state with quantum numbers $\alpha \ell j$ at energy $E$.
   The spectral strength for a given $\ell j$ combination can be found by summing (integrating) the spectral function according to
   \begin{equation}
      S_{\ell j}(E) = \sum_\alpha S_{\ell j}(\alpha,\alpha;E) .
      \label{eq:strength}
   \end{equation}
   The spectral strength $S_{\ell j}(E)$ is the contribution at energy $E$ to the occupation from all orbitals with $\ell j$.
The occupation of specific orbits characterized by $n$ with wave functions that are normalized to 1 can be obtained from Eq.~(\ref{eq:spec}) by folding in the corresponding wave functions~\cite{Dussan:2014},
   \begin{equation}
      S^{n-}_{\ell j}(E) = \sum_{\alpha,\beta}[\phi^n_{\ell j}(\alpha)]^*S^h_{\ell j}(\alpha,\beta;E)\phi^n_{\ell j}(\beta) .
      \label{eq:qh_strength}
   \end{equation}
Note that this representation of the spectral strength involves off-diagonal elements of the propagator.

Of particular interest are the solutions of the Dyson equation that correspond to discrete bound states with one proton removed. In the IPSM, these correspond to the $0\textrm{d}\frac{3}{2}$ and $1\textrm{s}\frac{1}{2}$ orbits for which $(e,e'p)$ cross sections are available and discussed in this paper.
Such quasihole wave functions are obtained from the nonlocal Schr\"{o}dinger-like equation 
   \begin{eqnarray}
      \sum_\gamma\bra{\alpha}T_{\ell} + \Sigma^*_{\ell j}(E)\ket{\gamma}\psi_{\ell j}^n(\gamma) = \varepsilon_n^-\psi_{\ell j}^n(\alpha),
      \label{eq:schrodinger}
   \end{eqnarray}
   where $\bra{\alpha}T_\ell\ket{\gamma}$ is the kinetic-energy matrix element, including the centrifugal term.
These wave functions correspond to overlap functions
   \begin{equation}
      \psi^n_{\ell j}(\alpha) = \bra{\Psi_n^{A-1}}a_{\alpha \ell j}\ket{\Psi_0^A}, \qquad \varepsilon_n^- = E_0^A - E_n^{A-1}.
      \label{eq:wavefunction}
   \end{equation}
Such discrete solutions to Eq.~(\ref{eq:wavefunction}) exist where there is no imaginary part of the self-energy, so near the Fermi energy. 
   The normalization for these wave functions is the spectroscopic factor, which is given by~\cite{Exposed!}
   \begin{equation}
      \mathcal{Z}^n_{\ell j} = \bigg(1 - \frac{\partial\Sigma_{\ell j}^*(\alpha_{qh},\alpha_{qh};E)}{\partial E}\bigg|_{\varepsilon_n^-}\bigg)^{-1},
      \label{eq:sf}
   \end{equation}
   where $\alpha_{qh}$ corresponds to the quasihole state that solves Eq.~(\ref{eq:schrodinger}). This corresponds to the spectral strength at the quasihole energy $\varepsilon_n^-$, represented by a delta function. Note that because of the presence of imaginary parts of the self-energy at other energies, there is also strength located there, thus the spectroscopic factor will be less than 1 and also less than the occupation probability.
   Indeed as shown in Ref.~\cite{Dussan:2014}, an equivalent spectral density  $S^p_{\ell j}(\alpha,\beta;E)$ for energies above $\varepsilon_F$ can be obtained which allows for the calculation of the presence of orbits that describe localized (and therefore normalized) single-particle states according to
   \begin{equation}
      S^{n+}_{\ell j}(E) = \sum_{\alpha,\beta}[\phi^n_{\ell j}(\alpha)]^*S^p_{\ell j}(\alpha,\beta;E)\phi^n_{\ell j}(\beta) .
      \label{eq:qp_strength}
   \end{equation}

The distribution of single-particle strength for the two relevant proton orbits will be discussed in Sec.~\ref{sec:results}.
 It reveals that the strength for these orbits is fragmented over all energies, positive and negative, rather than 
   concentrated at one energy as in the IPSM.
We note that the distribution at positive energies is constrained by elastic-scattering data, making the conclusion of the relevance of correlations beyond the IPSM inevitable~\cite{Dussan:2014}.
The strength of each orbit is peaked at its quasihole energy $\varepsilon_n^-$. 
The spectral strength distribution below $\varepsilon_F$ is constrained by the charge density and particle number which also receive contributions from other $\ell j$ quantum numbers~\cite{Exposed!}.

It is appropriate to introduce the Fermi energies for removal and addition given by
\begin{equation}
\label{eq:5.11a}
\varepsilon^-_F = E^A_0 - E^{A-1}_0
\end{equation}
and
\begin{equation}
\varepsilon^+_F = E^{A+1}_0 - E^A_0 ,
\label{eq:5.11b}
\end{equation}
referring to the ground states in the $A\pm1$ systems, respectively.
It is also convenient to employ the average Fermi energy
\begin{equation}
\varepsilon_F \equiv \frac{1}{2} \left[
\varepsilon_F^+  - \varepsilon_F^- \right]  .
\label{eq:FE}
\end{equation}
In practical work, we adhere to the average Fermi energy to separate the particle and hole domain and their corresponding imaginary parts of the self-energy. 
For specific questions related to valence holes, the imaginary part of the self-energy can be neglected and Eqs.~(\ref{eq:schrodinger}) and (\ref{eq:sf}) can be applied.
   The occupation probability of each orbital is calculated by integrating all contributions from the spectral strength up to the Fermi energy
   \begin{equation}
      n^n_{\ell j} = \int_{-\infty}^{\varepsilon_F} \!\!\!\! dE\ S^{n-}_{\ell j}(E) ,
      \label{eq:occ}
   \end{equation}
   whereas the depletion of the orbit is obtained from
      \begin{equation}
      d^n_{\ell j} = \int_{\varepsilon_F}^\infty \!\!\!\! dE\ S^{n+}_{\ell j}(E) .
      \label{eq:dep}
   \end{equation}
   Since the DOM has so far been limited to 200 MeV positive energy, a few percent of the sum rule
   \begin{equation}
   n^n_{\ell j} + d^n_{\ell j} = 1 ,
   \label{eq:antic}
   \end{equation}
  that reflects the anticommutator relation of the corresponding fermion addition and removal operators, 
    has been found above this energy~\cite{Dussan:2014}.
   The particle number of the nucleus is found by summing over each $\ell j$ combination while integrating the spectral strength up to the Fermi energy, 
   \begin{equation}
      Z,N = \sum_{\ell j} (2j+1) \int_{-\infty}^{\varepsilon_F} \!\!\!\! dE\  S^{p,n}_{\ell j}(E) .
   \end{equation}
   where $Z$ and $N$ are the total number of protons and neutrons, respectively. 
   The DOM calculation of $^{40}$Ca that includes $\ell \le 5$, results in $Z=19.8$ and $N=19.7$.
   As 20 is the experimental number, this allows for small contributions from higher $\ell$-values.

   \subsection{Dispersive optical model}
\label{sec:DOM}
   It was recognized long ago that the irreducible self-energy represents the potential 
   that describes elastic-scattering observables~\cite{Bell59}. 
   The link with the potential at negative energy is then provided by the Green's function framework as was realized by Mahaux and Sartor who introduced the DOM as reviewed in Ref.~\cite{Mahaux91}. 
   The analytic structure of the nucleon self-energy allows one 
to apply the dispersion relation, which relates the real part of the self-energy at a given energy to a dispersion integral of its imaginary part over all energies.
   The energy-independent correlated Hartree-Fock (HF) contribution~\cite{Exposed!} is removed by employing a subtracted dispersion relation with the Fermi energy used as the subtraction point~\cite{Mahaux91}.
   The subtracted form has the further advantage that the emphasis is placed on energies closer to the Fermi energy for which more experimental data are available.
   The real part of the self-energy at the Fermi energy is then still referred to as the HF term, but is sufficiently attractive to bind the relevant levels.
   In practice, the imaginary part is assumed to extend to the Fermi energy on both sides while being very small in its  vicinity.
The subtracted form of the dispersion relation employed in this work is given by
   \begin{eqnarray}
      \textrm{Re}\ \Sigma^*(\alpha,\beta;E) = \textrm{Re}\ \Sigma^*(\alpha,\beta;\varepsilon_F) \qquad\qquad\qquad\qquad \label{eq:dispersion} \\* 
      -  \mathcal{P}\int_{\varepsilon_F}^{\infty} \!\! \frac{dE'}{\pi}\textrm{Im}\ \Sigma^*(\alpha,\beta;E')[\frac{1}{E-E'}-\frac{1}{\varepsilon_F-E'}] \nonumber \\* 
      + \mathcal{P} \! \int_{-\infty}^{\varepsilon_F} \!\! \frac{dE'}{\pi}\textrm{Im}\ \Sigma^*(\alpha,\beta;E')[\frac{1}{E-E'}-\frac{1}{\varepsilon_F-E'}], \nonumber      
   \end{eqnarray}
   where $\mathcal{P}$ is the principal value. 
   The static term is denoted by  $\Sigma_{\text{HF}}$ from here on. 
   Equation~(\ref{eq:dispersion}) constrains the real part of the self-energy through empirical information of the HF term and empirical knowledge of the imaginary part, which is closely tied to experimental data. 
   Initially, standard functional forms for these terms were introduced by Mahaux and Sartor who also cast the DOM potential in a local form by a standard transformation which turns a nonlocal static HF potential into an energy-dependent local potential~\cite{Perey:1962}.
   Such an analysis was extended in Refs.~\cite{Charity06,Charity:2007} to a sequence of Ca isotopes and in Ref.~\cite{Mueller:2011} to semi-closed-shell nuclei heavier than Ca.
   The transformation to the exclusive use of local potentials precludes a proper calculation of nucleon particle number and expectation values of the one-body operators, like the charge density in the ground state. 
   This obstacle was eliminated in Ref.~\cite{Dickhoff:2010}, but it was shown that the introduction of nonlocality in the imaginary part was still necessary in order to accurately account for particle number and the charge density~\cite{Mahzoon:2014}.
   Theoretical work provided further support for this introduction of a nonlocal representation of the imaginary part of the self-energy~\cite{Waldecker:2011,Dussan:2011}.
   A recent review has been published in Ref.~\cite{Dickhoff:2017}.

   We implement a nonlocal representation of the self-energy following Ref.~\cite{Mahzoon:2014} where $\Sigma_{\text{HF}}(\bm{r},\bm{r'})$ and $\textrm{Im}\ \Sigma(\bm{r},\bm{r'};E)$ are parametrized, using Eq.~(\ref{eq:dispersion}) to generate the energy dependence of the
   real part. The HF term consists of a volume term, spin-orbit term,  and a wine bottle shape~\cite{Brida11} to simulate a surface contribution. The imaginary self-energy consists of volume, surface, and spin-orbit terms. 
   Details can be found in~\cite{Mahzoon:2014}. Nonlocality is represented using the Gaussian form 
   \begin{equation}
      H(\bm{s},\beta) = \pi^{-3/2}\beta^{-3}e^{-\bm{s}^2/\beta^2} ,
      \label{eq:nonlocality}
   \end{equation}
   where $\bm{s} = \bm{r} -\bm{r}'$, 
   as proposed in Ref.~\cite{Perey:1962}. 
  As mentioned previously, it was customary in the past to replace nonlocal potentials by local, energy-dependent potentials~\cite{Mahaux91,Perey:1962,Fiedeldey:1966,Exposed!}. The introduction of an energy dependence alters the dispersive
   correction from Eq.~(\ref{eq:dispersion}) and distorts the normalization, leading to incorrect spectral functions and related quantities~\cite{Dickhoff:2010}. Thus, a nonlocal implementation permits the self-energy to accurately 
   reproduce important observables such as the charge density and particle number. 
   Only the nonlocal version of the DOM is therefore particularly well suited for describing $(e,e'p)$ cross sections.

   In order to use the DOM self-energy for predictions, the parameters are fit through a weighted $\chi^2$ minimization of available elastic differential cross section data ($\frac{d\sigma}{d\Omega}$), analyzing power data ($A_\theta$),  reaction cross sections ($\sigma_r$), total cross sections ($\sigma_t$), charge density ($\rho_{\text{ch}}$), energy levels ($\varepsilon_{\ell j}$), particle number, separation energies, 
   and root-mean-square charge radius ($r_{\text{rms}}$).
   The potential is transformed from coordinate-space to a Lagrange basis using Legendre and Laguerre polynomials for scattering and bound-states, respectively.
   The bound-states are found by diagonalizing the Hamiltonian in Eq.~(\ref{eq:schrodinger}), the propagator is found by inverting the Dyson equation, Eq.~(\ref{eq:dyson}), 
   while all scattering calculations are done in the framework of $R$-matrix theory~\cite{Baye:2010}. 
Predictions of the DOM have been published in Ref.~\cite{Mahzoon:2017} where a large neutron skin for ${}^{48}$Ca was generated.

   \subsection{DWIA description of the $(e,e'p)$ cross section}
\label{sec:DWIA}
In the past, $(e,e'p)$ cross sections obtained at Nikhef in Amsterdam have been successfully described by utilizing the DWIA.
This description is expected to be particularly good when kinematics is used that emphasizes the longitudinal coupling of the excitation operator, which is dominated by a one-body operator.
The Nikhef group was able to fulfill this condition by choosing kinematical conditions in which the removed proton carried momentum parallel or antiparallel to the momentum of the virtual photon.
Under these conditions, the transverse contribution involving the spin and possible two-body currents is suppressed. Therefore the process can be interpreted as requiring an accurate description of the transition amplitude connecting the resulting excited state to the ground state by a known one-body operator.
This transition amplitude is contained in the polarization propagator which can be analyzed with a many-body description involving linear response~\cite{Exposed!}.
Such an analysis demonstrates that the polarization propagator contains two contributions.
The first term involves the propagation of a particle and a hole dressed by their interaction with the medium, but not each other.
The other term involves their interaction. The latter term will dominate at low energy when the proton that absorbs the photon participates in collective excitations like surface modes and giant resonances. When the proton receives on the order of 100 MeV it is expected that the excited state that is created can be well approximated by the dressed particle and dressed hole excitation~\cite{Brand90}.
In fact, when strong transitions are considered, like in the present work, two-step processes have only minor influence~\cite{Kramer:1989,Gerard_12C}. 
This interpretation forms the basis of the DWIA applied to exclusive $(e,e'p)$ cross sections obtained by the Nikhef group.
The ingredients of the DWIA therefore require a proton distorted wave describing the outgoing proton at the appropriate energy and an overlap function with its normalization for the removed proton.
The distorted wave was typically obtained from a standard local global optical potential like Ref.~\cite{Schwandt:1982} for ${}^{40}$Ca.
The overlap function was obtained by adjusting the radius of a local Woods-Saxon potential to the shape of the $(e,e'p)$ cross section while adjusting its depth to the separation energy of the hole.
Its normalization was obtained by adjusting the calculated DWIA cross section to the actual data~\cite{Lapikas93}.  
Standard nonlocality corrections were applied to both the outgoing and removed proton wave functions~\cite{Perey:63}, in practice making the bound-state wave function the solution of a nonlocal potential.
We observe that such corrections are $\ell$-independent and therefore different from the nonlocal DOM implementation.

   In order to describe the $(e,e'p)$ reaction, the incoming electron, the electron-proton interaction, the outgoing electron, and the outgoing proton must therefore be addressed. 
   The cross section is calculated from the hadron tensor, $W^{\mu\nu}$,  which contains matrix elements of the nuclear charge-current density, $J^\mu$~\cite{ElectroResponse}.  
   Using the DWIA, which assumes that the virtual photon exchanged by the electron couples to the 
   same proton that is detected and the final-state interaction can be described using an optical potential~\cite{Giusti:1988,Boffi:1980}, the nuclear current can be written as
   \begin{equation}
      J^\mu(\bm{q}) = \int d\bm{r}e^{i\bm{q}\cdot\bm{r}}\chi^{(-)*}_{E\ell j}(\bm{r})(\hat{J}^\mu_{\text{eff}})_{E\ell j}(\bm{r})\psi^n_{\ell j}(\bm{r})\sqrt{\mathcal{Z}^n_{\ell j}},
      \label{eq:current}
   \end{equation}
   where $\chi^{(-)*}_{E}(\bm{r})$ is the outgoing proton distorted wave~\cite{ElectroResponse}, 
   $\psi^n_{\ell j}$ is the overlap function, $\mathcal{Z}^n_{\ell j}$ its normalization, $\bm{q} = \bm{k_f} - \bm{k_i}$ is the electron three-momentum transfer, and 
   $\hat{J}^\mu_{\text{eff}}$ is the effective current operator~\cite{ElectroResponse}. 
   The incoming and outgoing electron waves
   are treated within the Effective Momentum Approximation, where the waves are represented by plane waves with effective momenta to account for distortion from the interaction with the target 
   nucleus~\cite{Giusti:1987}
   \begin{equation}
      k_{i(f)}^{\text{eff}} = k_{i(f)} + \int d\bm{r}V_c(\bm{r})\phi_{\ell j}^2(\bm{r}),
      \label{eq:effective}
   \end{equation}
   where $V_c(\bm{r})$ is the Coulomb interaction.
   This alters Eq.~(\ref{eq:current}) by replacing $\bm{q}$ with $\bm{q}_{\text{eff}}$.

   In the plane-wave impulse approximation (PWIA), in which the outgoing proton wave is approximated by a plane wave, the $(e,e'p)$ can be factorized into an off-shell electron-proton cross section and the spectral function~\cite{ElectroResponse}, 
   \begin{equation}
      S(E_m,\bm{p}_m) = \frac{1}{k\sigma_{ep}}\frac{d^6\sigma}{dE_{e'}d\Omega_{e'}dE_pd\Omega_p}.
      \label{eq:momdist}
   \end{equation}
   The off-shell electron-proton cross section, $\sigma_{ep}$, is approximated from the on-shell one using the $\sigma_{\text{cc1}}$ model as proposed in~\cite{deForest:1983}.
  This separation does not hold true for the DWIA, but the displayed cross sections, both the experimental and theoretical ones, have been divided by the $\sigma_{\text{cc1}}$ cross section.
  Note that Eq.~(\ref{eq:momdist}) is equivalent to the diagonal element of Eq.~(\ref{eq:spec}) when the momentum basis is employed and the restriction to given values of $\ell$ and $j$ is taken into account.
  In principle, corrections due to two-step processes could be considered but they are estimated to make negligible contributions for the transitions considered in this study~\cite{Kramer:1989}.
   
The calculations of the $(e,e'p)$ cross sections in this paper were performed by employing DOM ingredients that were constrained by other experimental data. Appropriate distorted waves and overlap functions with their normalization were thus generated that allow for a DWIA description of the exclusive $(e,e'p)$ cross section for valence holes in ${}^{40}$Ca. Agreement with cross sections therefore supports the description of the reaction in a DWIA framework, but also confirms the overall consistency of the DOM approach including its interpretation of the normalization of the overlap functions as spectroscopic factors that can be confronted with data.
   
\section{Experiment}
\label{sec:EXP}

\begin{figure}[t]
   \begin{minipage}{\columnwidth}
      \makebox[\columnwidth]{
         \includegraphics[scale=0.72]{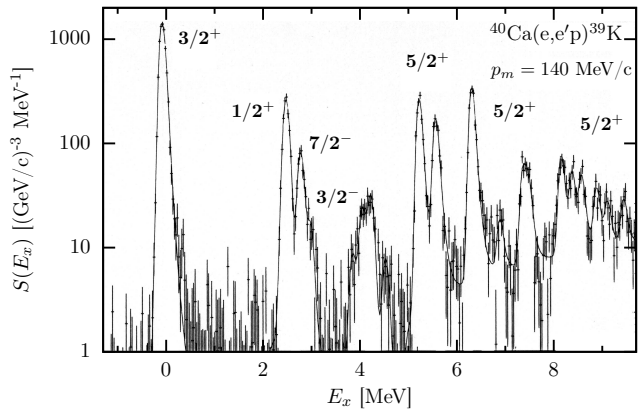}
      }
   \end{minipage}
   \caption{
   Radiatively unfolded excitation-energy spectrum for the reaction
${}^{40}$Ca$(e,e'p)$ at missing momentum 140 MeV/c, showing the well resolved transitions to
the $J^\pi = 3/2^+$ ground state and $1/2^+$ first excited state in ${}^{39}$K. 
Above $E_x$ = 5 MeV several transitions to states with mostly $J^\pi = 5/2^+$ are identified.
The peak at $E_x \approx 4$ MeV results from the reaction ${}^{16}$O$(e,e'p)^{15}\textrm{N}_{g.s.}$ due to oxygen contamination in the target. The curve is a multiple gaussian fit to the data.
 }
   \label{fig:excite}
\end{figure}

   The experimental data for the reaction $^{40}$Ca$(e,e'p)^{39}$K that are presented in this paper were obtained with the  electron beam from the Medium Energy Accelerator (MEA) at Nikhef in Amsterdam with natural calcium targets ($^{40}$Ca abundance about 97\%) of thicknesses 14.3 and 24.6 mg/cm$^2$. Typical values for the beam current amounted to several $\mu A$, while the duty factor of the beam was about 1\%. The beam energies $E_0$ used were between 299 and 532 MeV, as listed in Table~\ref{tab:expt}. The beam was tuned in so called ``double dispersion matching'' mode~\cite{LapW80}, which resulted in a missing-energy resolution in the range 130-200 keV (see Fig.~\ref{fig:excite}).

The experiment was carried out in the EMIN hall~\cite{Vri84}, where the scattered electrons and ejected protons were detected in a pair of high-resolution magnetic spectrometers~\cite{Vri84}. Each of the spectrometers had a momentum acceptance of $\pm$ 5\% 
   around the central value. The luminosity determination was calibrated by comparing measured elastic electron scattering from $^{12}$C to known literature values~\cite{deVries:1987}. 

The amount of $^{40}$Ca in the targets was determined with an accuracy of 2\% from a comparison of  elastic electron scattering from $^{40}$Ca to known literature values~\cite{deVries:1987}. Since the targets also contained some oxygen and hydrogen (weight less than 11\%), bound as Ca(OH)$_2$, the missing-energy spectra showed peaks for the reactions $^{16}$O$(e,e'p)$$^{15}$N$_{g.s.}$ and $^1$H$(e,e'p)$. Due to the excellent missing-energy resolution, these peaks were resolved, and, moreover, are outside the missing-energy range of interest in the present analysis.
 An example of the quality of the data is displayed in Fig.~\ref{fig:excite}, demonstrating the fragmentation of the strength for $T_p$=100 MeV and $p_m$=140 MeV/c. Different spin-parity identifications are displayed when known from other experiments.

The efficiency of coincidence detection $\epsilon_c$ was obtained from a comparison of the overcomplete coincidence reaction $^1$H$(e,e'p)$ with a simultaneous measurement of singles elastic scattering $^1$H$(e,e')$ (see Table~\ref{tab:expt}).

   The data were collected in so called ``parallel kinematics'', in which the momentum $\bm{p}$ of the ejected proton is in the same direction as the momentum transfer $\bm{q}$ of the virtual photon. 
   Since the cross section for the reaction depends sensitively on the energy of the ejected proton, we measured three sets of data at proton kinetic energies $T_p$ = 70, 100 and 135 MeV, respectively (see Table~\ref{tab:expt}). 
   \begin{table}
      \caption{Survey of experimental parameters: central ejected proton energy $T_p$, range of employed electron beam energies $E_0$, measured coincidence detection efficiency $\epsilon_c$, and total systematic error $\Delta\sigma/\sigma_{syst}$.}
      \label{tab:expt}
      \begin{tabular}{lcccccc}
         \\
         \hline
         &  $T_p$   & $E_0$  & $\epsilon_c$ & $\Delta\sigma/\sigma_{syst}$\\
         \hline
         & MeV   & MeV  & \%         &  \%  \\
         \hline
         A& 70 & 299 - 483 & 97.1$\pm$ 1.1  & 6.0 \\
         B& 100  & 313 - 532 & 98.5$\pm$ 0.5 & 2.8 \\
         C&  135  & 483     & 97.1$\pm$ 1.1  & 6.0 \\
         \hline
      \end{tabular}
   \end{table}

   From the measured coincidence events the experimental six-dimensional differential cross sections were determined in the standard way described extensively in Ref.~\cite{Denherder:1988}.
   These coincidence cross sections were subsequently unfolded for radiative effects according to the method described in Ref.~\cite{Quint88} and then converted to reduced cross sections using Eq.~(\ref{eq:momdist}). 

   The $T_p$=100 MeV data were analyzed previously in Ref.~\cite{Kramer:1989} with bound-state wave functions calculated in a Woods-Saxon well (free parameters:~well radius and spectroscopic factor) and distorted outgoing proton wave functions calculated in a global energy-dependent optical-model potential described by Schwandt \textit{et al.}~\cite{Schwandt:1982}. A study of the mechanism of the reaction $(e,e'p)$, including the present $^{40}$Ca data, was published earlier by one of us~\cite{Blok93}.
  
  In the present paper the experimental data are compared at three proton energies to predictions of the DWIA using only DOM ingredients discussed in Sec.~\ref{sec:DOM}.
  For this purpose, the well-resolved transitions to the ground state ($3/2^+$) and first excited state ($1/2^+$) at 2.522 MeV in $^{39}$K were selected. 
  In order to facilitate the comparison, the reduced cross sections $\sigma^{exp}(p_m,E^0_i,\theta_i)$ in each data set A, B, and C were transformed to the highest-employed beam energy $E^0_h$ in that set according to
   \begin{equation}
      \sigma^{exp}_{tr} (p_m,E^0_h,\theta_h) = \frac {\sigma^{th}(p_m,E^0_h,\theta_h)} {\sigma^{th}(p_m,E^0_i,\theta_i)} {\sigma^{exp}(p_m,E^0_i,\theta_i)},
      \label{eq:transform}
   \end{equation}
 where in parallel kinematics the scattered electron angle $\theta_h$ follows directly from momentum and energy conservation given the fixed value of $T_p$.
   The model dependence of such a transformation was found to be less than 1\%, as derived from a comparison of the transformed cross section obtained with the Schwandt optical potential~\cite{Schwandt:1982} ($th$=Schwandt) and the present DOM potential ($th$=DOM), respectively. 
   In the final step, experimental momentum distributions for the two transitions were determined by integration of the transformed reduced cross sections over the  missing-energy region covering the corresponding peak.
   The results are compared to theory in Sec.~\ref{sec:results}.

   \section{Results}
\label{sec:results}
\begin{figure}[t]
   \begin{minipage}{\columnwidth}
      \makebox[\columnwidth]{
         \includegraphics[scale=0.7]{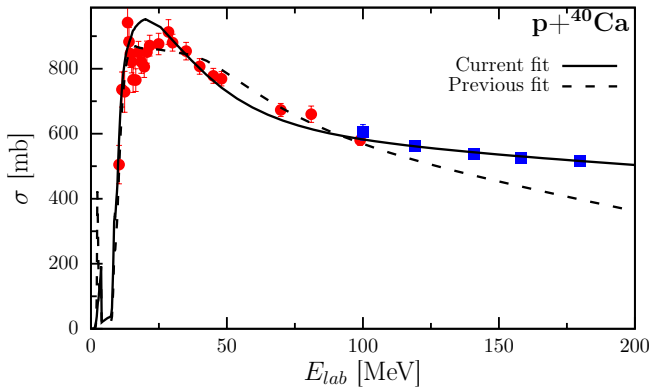}
      }
   \end{minipage}
   \caption{The proton reaction cross section for $^{40}$Ca. The solid line represents the current fit, while the dashed line depicts the previous fit~\cite{Mahzoon:2014}. The circular data points were included in the previous fit, while the square data points~\cite{PhysRevC.71.064606} have been added in the current fit.}
   \label{fig:react}
\end{figure}

The nonlocal DOM description of ${}^{40}$Ca data was presented in Ref.~\cite{Mahzoon:2014}.
In the mean time, additional experimental higher-energy proton reaction cross sections~\cite{PhysRevC.71.064606} have been incorporated which caused some adjustments of the DOM parameters compared to Ref.~\cite{Mahzoon:2014}. The updated parameters are collected in Appendix A.
Adjusting the parameters from the previous values~\cite{Mahzoon:2014} to describe these additional experimental results leads to an equivalent description for all data except these reaction cross sections.
   These higher-energy data dictate that the proton reaction cross section stay flat for energies in the region around 150 MeV, as shown in Fig.~\ref{fig:react}. This means there is more absorption at higher energies than in the previous fit, leading to increased strength in the imaginary part of the self-energy. Due to the dispersion relation, Eq.~(\ref{eq:dispersion}), this increases the spectral strength at positive energies when the Dyson equation is solved. The sum rule pertaining to the integral over all energies of the strength of the valence holes then implies that strength is transfered from below the Fermi energy to the energies with an increased imaginary part. 
   This resulting loss of strength below the Fermi energy reduces the spectroscopic factors by about 0.05 compared to the results reported in Ref.~\cite{Mahzoon:2014}.

\begin{figure}[b]
      \begin{minipage}{\linewidth}
         \makebox[\linewidth]{
            \includegraphics[scale=0.7]{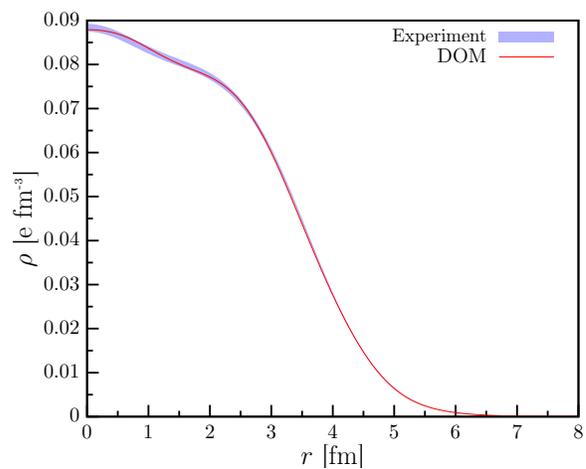}
         }
      \end{minipage}
      \caption{Experimental and fitted $^{40}$Ca charge density. The solid line is calculated using the DOM propagator, 
      while the experimental band represents the 1\% error associated with the extracted charge density from elastic electron scattering experiments~\cite{deVries:1987,Sick79}.}
   \label{fig:chd}
\end{figure}

To accurately calculate the $(e,e'p)$ cross section in DWIA, it is imperative that the DOM self-energy describe not only scattering data but bound-state information as well. 
This is due to the fact that the shape of the cross section is primarily determined by the bound-state overlap function~\cite{Kramer:1989}. Thus, not only should the experimental charge radius be reproduced, but the charge density should match the experimental data, as we report in Fig.~\ref{fig:chd}, where the DOM charge density is shown as the solid line and compared with the deduced charge density obtained from~\cite{deVries:1987} with the band representing the 1\% error.  
We employed the Fourier-Bessel parametrization~\cite{deVries:1987} that accurately reproduces the data reported in Ref.~\cite{Sick79}.

\begin{figure}[t]
   \begin{minipage}{\columnwidth}
      \makebox[\columnwidth]{
         \includegraphics[scale=0.7]{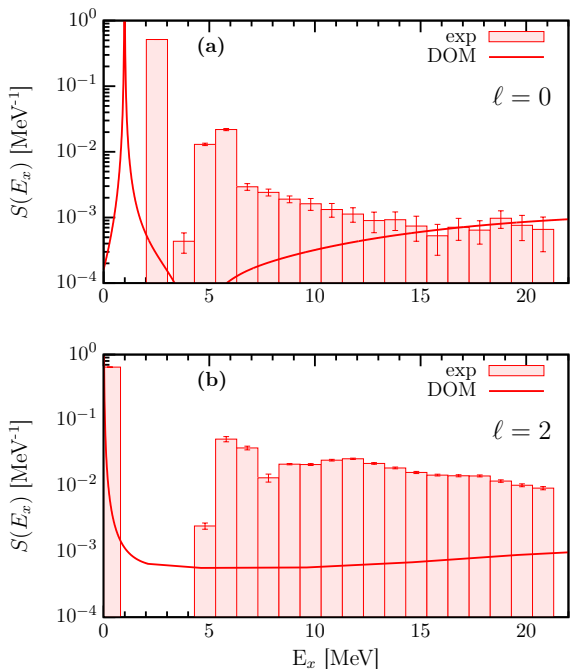}
      }
   \end{minipage}
   \caption{ Spectral strength as a function of excitation energy for (a) the $1\textrm{s}\frac{1}{2}$ and (b) the $0\textrm{d}\frac{3}{2}$ proton orbitals, calculated from the DOM using Eq.~(\ref{eq:strength}) 
   and extracted from the $^{40}$Ca$(e,e'p)$$^{39}$K experiment~\cite{Kramer:1989,Kramer90}. 
   The  peaks in the DOM curves and experimental data correspond to the quasihole energies of the protons in $^{40}$Ca. The DOM peak in (a) does not exactly match the experiment (see Ref.~\cite{Mahzoon:2014}).
   The distance between the quasihole peak and the smaller contributions is substantially larger in (b) than in (a).
    Note that the experimental fragments in (b) above 4 MeV mostly correspond to $0\textrm{d}\frac{5}{2}$ strength.} 
   \label{fig:spectral}
\end{figure} 
   The present DOM self-energy leads to the spectral strength distributions in Fig.~\ref{fig:spectral}. 
   The experimental points are the results of an angular-momentum decomposition of the experimental spectral function at $T_p$ = 100 MeV as described in Ref.~\cite{Kramer90}.
   The experimental distributions for $\ell = 0,2$ clearly show that the 
   strength is already strongly fragmented at low energies. 
   The main peak in each case represents the valence hole transition of interest in this paper.
   The DOM strength is plotted as a continuous function employing the imaginary part of the self-energy, which is very small near the Fermi energy, to clarify that only one peak is generated in the present implementation.
   The DOM therefore does not yet include the details of the low-energy fragmentation of the valence hole states which requires the introduction of pole structure in the self-energy~\cite{Dickhoff04}.
   The spectroscopic factor of Eq.~(\ref{eq:sf}) corresponds to the main peak of each distribution shown in Fig.~\ref{fig:spectral}. 
   It is calculated directly from the ${}^{40}$Ca DOM self-energy resulting in values of 0.71 and 0.74 for the $0\textrm{d}\frac{3}{2}$ and $1\textrm{s}\frac{1}{2}$ peaks, respectively.
The results are probed in more detail by analyzing the momentum distributions of the $^{40}$Ca$(e,e'p)$$^{39}$K reaction.
 
\begin{figure}[t]
      \begin{minipage}{\linewidth}
         \makebox[\linewidth]{
            \includegraphics[scale=0.7]{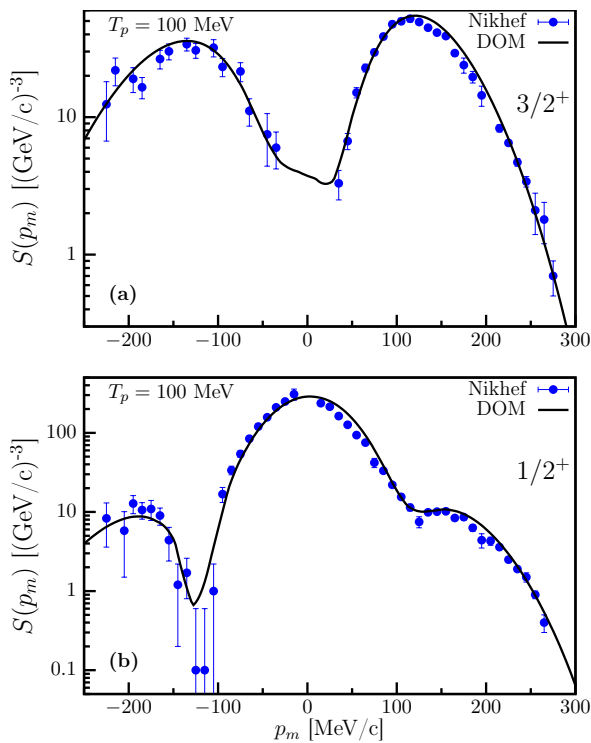}
         }
      \end{minipage}
      \caption{$^{40}$Ca$(e,e'p)$$^{39}$K spectral functions in parallel kinematics at an outgoing proton kinetic energy of 100 MeV. The solid line is the calculation using the DOM ingredients, while the points are from the 
      experiment detailed in~\cite{Kramer:1989}.
      (a) Distribution for the removal of the $0\textrm{d}\frac{3}{2}$. The curve contains the DWIA for the $3/2^+$ ground state including a spectroscopic factor of 0.71. (b) Distribution for the removal of the $1\textrm{s}\frac{1}{2}$ proton with a spectroscopic factor of 0.60 for the $1/2^+$ excited state at 2.522 MeV.}
   \label{fig:eep100}
\end{figure}

\begin{figure}[t]
      \begin{minipage}{\linewidth}
         \makebox[\linewidth]{
            \includegraphics[scale=0.7]{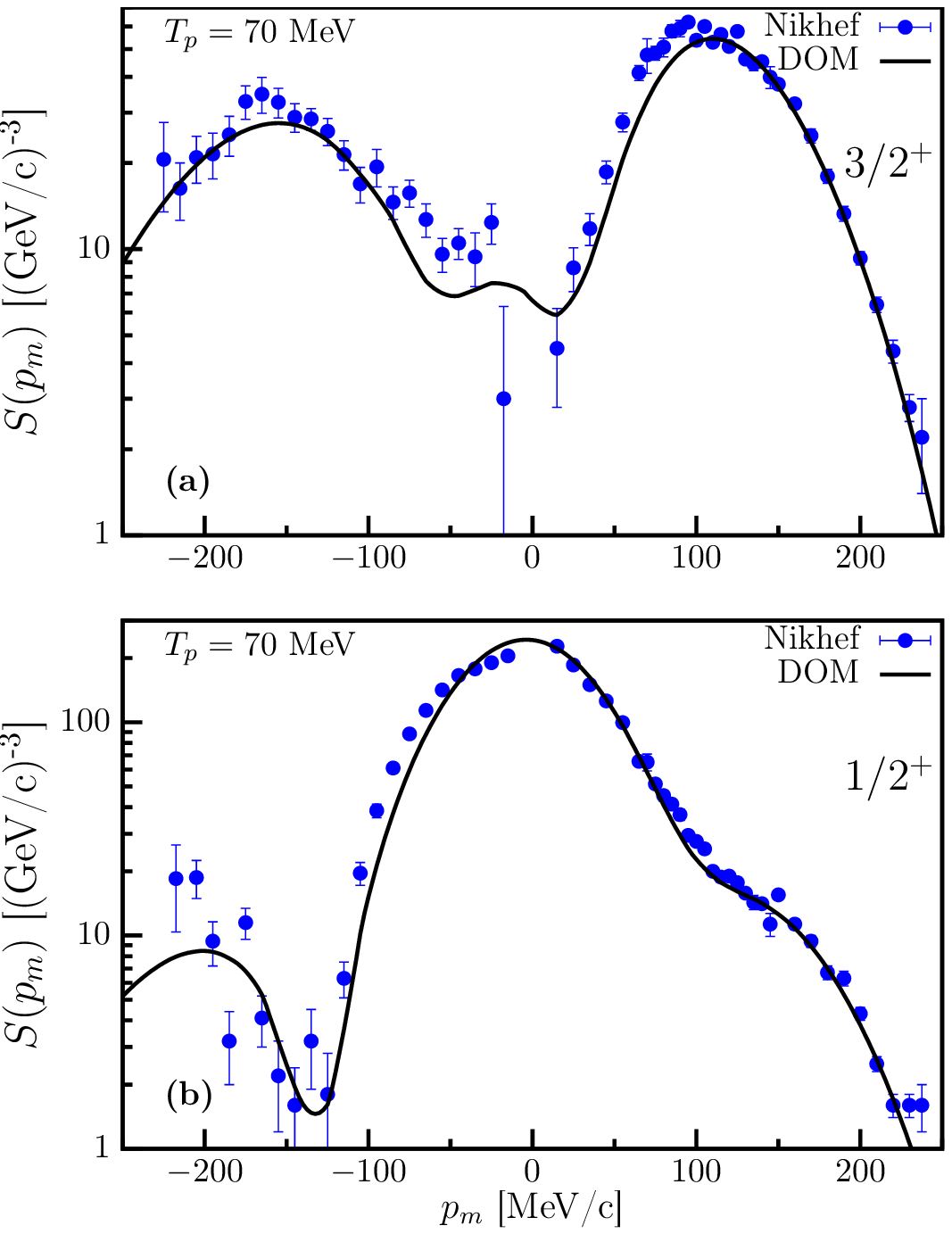}
         }
      \end{minipage}
      \caption{As for Fig.~\ref{fig:eep100}, but for an outgoing proton energy of 70 MeV.}
   \label{fig:eep70}
\end{figure}

\begin{figure}[t]
      \begin{minipage}{\linewidth}
         \makebox[\linewidth]{
            \includegraphics[scale=0.7]{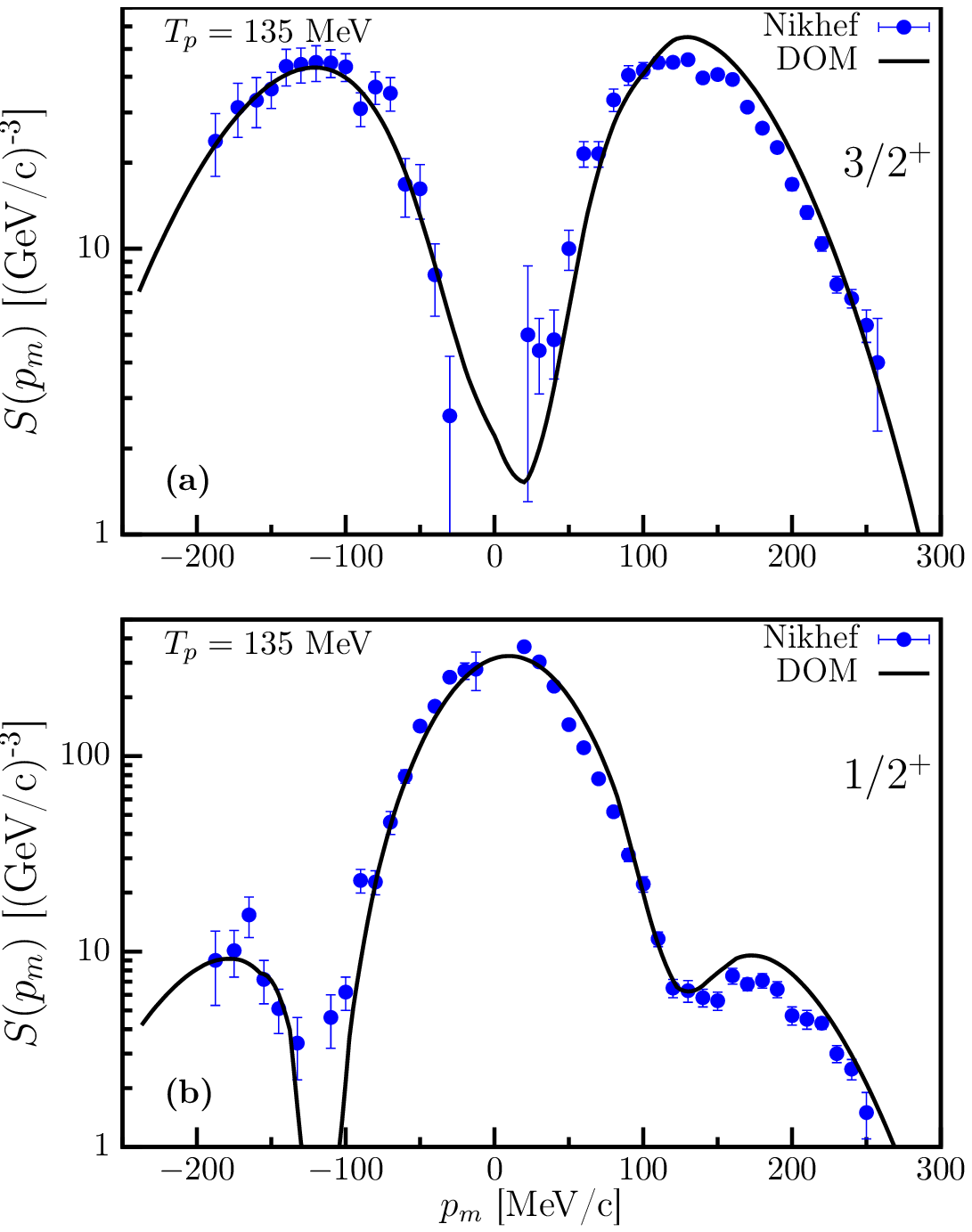}
         }
      \end{minipage}
      \caption{As for Fig.~\ref{fig:eep100}, but for an outgoing proton energy of 135 MeV.}
   \label{fig:eep135}
\end{figure}

   In the past, the DWIA calculations by the Nikhef group have been performed using the DWEEPY code~\cite{Giusti:1988}.
   For the present work the momentum distributions are calculated by adapting a recent version of the DWEEPY code~\cite{Giusti:2011} to use the DOM bound-states, distorted waves, and spectroscopic factors as inputs. 
   Before confronting the DOM calculations with the experimental cross sections it is necessary to consider the consequences of the low-energy fragmentation as shown in Fig.~\ref{fig:spectral}.
   For the $0\textrm{d}\frac{3}{2}$ ground state transition there is a clear separation with higher-lying fragments, most of which cannot be distinguished from $0\textrm{d}{\frac{5}{2}}$ contributions as the experiments were not able to provide the necessary polarization information.
In addition, these higher-lying fragments appear to carry little $0\textrm{d}\frac{3}{2}$ strength~\cite{Kramer:2001} , so the DOM spectroscopic factor can therefore be directly used to calculate the cross section of the ground-state peak.
The situation is different for the $1\textrm{s}\frac{1}{2}$ distribution which, while dominated by the large fragment at 2.522 MeV, exhibits substantial nearby strength as shown in Fig.~(\ref{fig:spectral})a.
   These contributions come from other discrete poles in the propagator, reflecting the mixing of the $1\textrm{s}\frac{1}{2}$ orbit to more complicated excitations nearby in energy. 
   Currently the origin of these additional discrete poles is not explicitly included in the DOM, although there is a smooth energy-dependent imaginary term in the self-energy to approximate their effect
   on the spectral strength~\cite{Exposed!}. 
   This approximation is sufficient when discussing integrated values such as the charge density and particle number, but falls short when considering details of the low-energy fragmentation into discrete energies as in the present situation. 
   The calculated DOM spectroscopic factor therefore includes strength in the neighborhood of the quasihole energy, resulting in an inflated value. 
   This effect is only noticeable in the $\ell=0$ case because there is a non-negligible amount of strength in the region near the peak.
   We turn to experimental data to account for this effect by enforcing that the ratio between the strength of the peak to the total spectral strength shown in the energy domain of Fig.~(\ref{fig:spectral}) 
   is the same between the data as for the DOM, 
   \begin{equation}
      \frac{\mathcal{Z}_F^{\text{DOM}}}{\int dE\ S^{\text{DOM}}(E)} = \frac{\mathcal{Z}_F^{\text{exp}}}{\int dE\ S^{\text{exp}}(E)}.
      \label{eq:ratio}
   \end{equation}
   Accounting for the contributions to the momentum distribution from different energies by scaling the DOM spectroscopic factor is justified by observing that the shape of the momentum distribution calculated at 
   similar energies is identical, with the strength being the only difference~\cite{Kramer:1989}. 
   The scaling of the spectroscopic factor leads to a reduction from 0.74 to 0.60.
   As mentioned, no correction is needed for the $0\textrm{d}\frac{3}{2}$ spectroscopic factor.
The resulting momentum distributions are shown in Figs.~\ref{fig:eep100}-\ref{fig:eep135}. The previous analysis of the Nikhef group at $T_p=100$ MeV~\cite{Kramer:1989} produced a comparable reproduction of the data with somewhat smaller spectroscopic factors, as shown in Table~\ref{table}.

 In order to estimate the uncertainty for the DOM spectroscopic factors, we followed the bootstrap method from Ref.~\cite{Varner91} which was also employed in Ref.~\cite{Mahzoon:2017} to assess the uncertainty for the neutron skin in ${}^{48}$Ca.
New modified data sets were created from the original data by randomly renormalizing each angular distribution or excitation function within the experimental error to incorporate fluctuations from the systematic errors. Twenty such modified data sets were generated and refit. The resulting uncertainties are listed in Table~\ref{table}.

  \begin{table}[b]
   \caption{Comparison of spectroscopic factors deduced from the previous analysis~\cite{Kramer:1989} using the Schwandt optical potential~\cite{Schwandt:1982} to the normalization of the corresponding overlap functions obtained in the present analysis from the DOM including an error estimate as described in the text.}
   \vspace{0.5cm}
         \begin{tabular}{ c c c } 
            \hline
            $\mathcal{Z}$ & $0\textrm{d}\frac{3}{2}$ & $1\textrm{s}\frac{1}{2}$\\
            \hline
            \hline
            Ref.~\cite{Kramer:1989} & $0.65 \pm 0.06$ & $0.51 \pm 0.05$\\
            \hline
            DOM & $0.71 \pm 0.04$ & $0.60 \pm 0.03$ \\
            \hline
         \end{tabular}
   \label{table} 
\end{table}

The DOM results yield at least as good agreement with the data as the standard analysis of Ref.~\cite{Kramer:1989} for the 100 MeV outgoing protons.
The main difference in the description can be pinpointed to the use of nonlocal potentials to describe the distorted waves.
Nonlocal potentials tend to somewhat suppress interior wave functions of scattering states and introduce an additional $\ell$ dependence as compared to local potentials.
We therefore conclude that the current consistent treatment clarifies that spectroscopic factors will be larger by about 0.05 when the proper nonlocal dispersive potentials are employed.
The DOM treatment of experimental data associated with both the particle and hole aspects of the single-particle propagator furthermore allows for a positive assessment of the quality of the DWIA to describe exclusive $(e,e'p)$ cross sections with outgoing proton energies around 100 MeV.


It is therefore fortunate that additional data have been obtained at 70 and 135 MeV to further delineate the domain of validity for the DWIA description of the reaction.
We document in Fig.~\ref{fig:eep70} the results when DOM ingredients are employed at this lower energy for the two valence hole states in ${}^{39}$K. 
The only difference in the DOM calculations for these cases is the use of a different proton energy, yielding different outgoing proton waves. The overlap function and the spectroscopic factors remain the same.
In Fig.~\ref{fig:eep70} the results are shown for $T_p = 70$ MeV. The description is of similar quality as the 100 MeV case.

 The agreement with the data at 135 MeV shown in Fig.~\ref{fig:eep135} is slightly worse but still acceptable. 
 At this energy (and corresponding value of the electron three-momentum transfer) the contribution of the transverse component of the excitation operator, where other mechanisms contribute in addition to those included in the present operator, will be larger. 
Given these results, it seems that parallel kinematics, in which the longitudinal part of the operator dominates, and a proton energy around 100 MeV, as chosen by the Nikhef group, is optimal for probing the removal probability of valence protons.
We note that this can only be achieved when an analysis is conducted in which all ingredients are provided by a nucleon self-energy that is constrained by all relevant available data as in the DOM.
The excellent agreement found here therefore supports the validity of the DOM approach as it is able to automatically account for the DWIA cross section in the domain where this approximation is expected to be valid.

   \begin{figure}[b]
      \begin{minipage}{\linewidth}
         \makebox[\linewidth]{
            \includegraphics[scale=0.7]{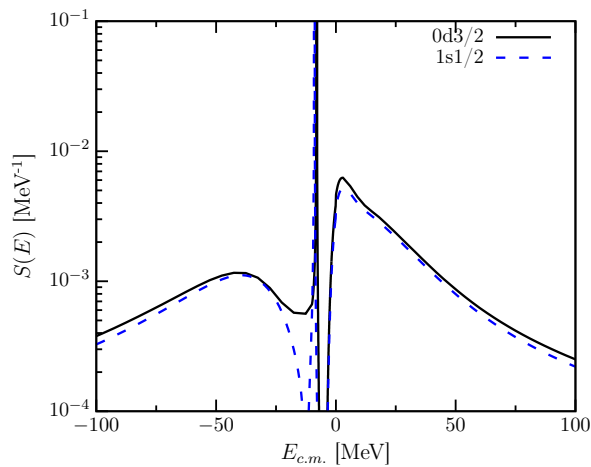}
         }
      \end{minipage}
      \caption{Spectral distribution of the $0\textrm{d}\frac{3}{2}$ and $1\textrm{s}\frac{1}{2}$ orbits as a function of energy. Additional strength outside this domain is not shown.}
   \label{fig:spectralsd}
\end{figure} 
The DOM results also generate the complete spectral distribution for the $0\textrm{d}\frac{3}{2}$ and $1\textrm{s}\frac{1}{2}$ orbits according to Eqs.~(\ref{eq:qh_strength}) and (\ref{eq:qp_strength}).
These distributions are displayed in Fig.~\ref{fig:spectralsd} from -100 to 100 MeV.
The energy axis refers to the $A-1$ system below the Fermi energy and the $A+1$ system above.
For plotting purposes the small imaginary part near the Fermi energy was employed giving the peaks a small width.
The occupation probabilities are obtained from Eq.~(\ref{eq:occ}) and correspond to 0.80 and 0.82 for the $0\textrm{d}\frac{3}{2}$ and $1\textrm{s}\frac{1}{2}$ orbits, respectively.
The strength at negative energy not residing in the DOM peak therefore corresponds to 9 and 7\%, respectively.
This information is constrained by the proton particle number and the charge density.
The strength above the Fermi energy is constrained by the elastic-scattering data and generates 0.17 and 0.15 for the $0\textrm{d}\frac{3}{2}$ and $1\textrm{s}\frac{1}{2}$ orbits, respectively, when Eq.~(\ref{eq:dep}) is employed up to 200 MeV.
The sum rule given by Eq.~(\ref{eq:antic}), associated with the anticommutation relation of the fermion operators, therefore suggests that an additional 3\% of the strength resides above 200 MeV, similar to what was found in Ref.~\cite{Dussan:2014}.
Strength above the energy where surface physics dominates can be ascribed to the effects of short-range and tensor correlations.
The main characterizations of the strength distribution shown in Fig.~55 of Ref.~\cite{Dickhoff04} are therefore confirmed for ${}^{40}$Ca.
The present results thus suggest that it is possible to generate a consistent picture of the strength distributions of these orbits employing all the available experimental constraints.
We therefore conclude that it is indeed quite meaningful to employ concepts like spectroscopic factors and occupation probabilities when discussing correlations in nuclei.

\section{Conclusions}
\label{sec:CON}
The main conclusion from the present work is that a consistent description of all available experimental data that are unambiguously related to the nucleon single-particle propagator is essential in providing accurate ingredients for a DWIA description of the $(e,e'p)$ reaction.
This description is provided by the DOM when it is implemented with nonlocal potentials up to at least 200 MeV in the elastic-scattering domain.
The availability of $(e,e'p)$ data at 70, 100, and 135 MeV of proton outgoing energy also delineates a window in which the DWIA provides an accurate description of the exclusive cross section with energies around 100 MeV appearing to be optimal.
We emphasize that it is also essential to consider the kinematical conditions that favor the longitudinal part of the excitation operator which is dominated by a one-body component.
This analysis therefore confirms the general conclusions reached in the past by the Nikhef group~\cite{Lapikas93}.

The confrontation of the DOM ingredients with the $(e,e'p)$ cross sections also demonstrates a necessary avenue for its further improvement.
It is fortunate that a rather complete experimental picture of the $\ell = 0$ fragmentation at low energy has also been determined utilizing the $(e,e'p)$ reaction~\cite{Kramer90}.
Using the experimental strength distribution without relying on their absolute values, it is possible to determine the fraction carried by the largest fragment at 2.522 MeV.
Since the DOM does not yet provide the details of this low-energy fragmentation, it was possible to identify the fraction of the DOM strength to be compared to the experimental cross section for the 2.522 MeV transition using this experimental information.
The resulting cross sections for both the ground state and 2.522 MeV state are then accurately described by the DWIA employing the DOM results.
Nevertheless, the DOM requires further improvement to incorporate more details on the low-energy fragmentation leading possibly to additional state dependence.
This improvement is particularly relevant for the description of strength distributions of weakly or deeply bound nucleons as they occur in $N-Z$ asymmetric nuclei.
Indeed, this feature must be addressed in the ongoing discussion related to spectroscopic factors deduced from transfer~\cite{Lee10} and knockout~\cite{Gade04} reactions, which appear to be in contradiction with each other. 
As has been highlighted here, it is important to clarify the amount of spectroscopic strength in the immediate vicinity of the main fragment.
This issue will only be more critical when a continuum of one of the nucleon species is nearby~\cite{Jensen11}.

The success of the DWIA for the description of the $(e,e'p)$ reaction has implications for the possibility of employing other reactions.
In particular, the $(p,pN)$ reaction above approximately 200 MeV incoming energy appears an attractive possibility~\cite{Ogata17}.
The availability of a proper description of the three distorted waves and the normalized overlap function using the DOM implies that it is possible to gauge the effective nucleon-nucleon interaction for this process by comparing with the $(e,e'p)$ results.
If successful, such an analysis would lend itself to an extension to rare isotopes for which this reaction is available~\cite{Aumann18,Kawase18}.
The current status of transfer reactions also suggests that the DOM can provide important contributions to the extraction of spectroscopic information~\cite{Nguyen11,Potel17}.
Before a consistent description of transfer reactions utilizing the DOM can be implemented, it will be necessary to improve the description of the deuteron distorted wave to the level currently achieved for single nucleons.

Finally, we can now shift the discussion of absolute spectroscopic factors to the level of observable $(e,e'p)$ cross sections in which the quality of the reaction description (DWIA) can be tested by a direct comparison with data.
Of particular value in reaching agreement with $(e,e'p)$ cross sections within the DOM framework is the availability of reaction cross section data, including those above 100 MeV, that directly quantify the strength of the coupling of the single-particle degree of freedom to other excitations through the imaginary part of the self-energy.
Our values for the valence spectroscopic factors of 0.71 for the $3/2^+$ ground state in ${}^{39}$K and 0.60 for the $1/2^+$ excited state appear to be the final answer in the quest for absolute values for ${}^{40}$Ca.
Taking into account the uncertainties associated with all the data that provide the presented self-consistent analysis, we employed a bootstrap error analysis to estimate that the accuracy in these values is $\pm0.04$.

\section*{Acknowledgement}
The authors thank  Carlotta Giusti for providing us with a recent version of the DWEEPY code.
This work was supported by the U.S. Department of Energy, Division of
Nuclear Physics under grant No. DE-FG02-87ER-40316, by the U.S. National Science Foundation under grant PHY-1613362, and the Dutch Foundation for Fundamenteel Onderzoek der Materie (FOM), which was financed by the Netherlands Organisation for Scientific Research (NWO). 

\appendix
\section{Description of the DOM potential}
\label{sec:appendix}
\subsection*{Parametrization}
\label{Sec:param}
We provide a detailed description of the parametrization of the proton and neutron self-energies in $^{40}$Ca used in the fits to bound and scattering data. The functional forms are the same as those from our previous study of $^{40}$Ca \cite{Mahzoon:2014} with the addition of a spin-orbit nonlocality. Including spin-orbit nonlocality requires an additional parameter, $\beta_{nl}$.
We use a simple Gaussian nonlocality in all instances \cite{Perey:1962}, corresponding to the HF term, real spin-orbit term, and to the volume and surface contributions to the imaginary part of the potential.
We write the HF self-energy term in the following form with the local Coulomb contribution.
\begin{eqnarray}
\Sigma_{HF}(\bm{r},\bm{r}') = \Sigma^{nl}_{HF}(\bm{r},\bm{r}') + V^{nl}_{so}(\bm{r},\bm{r}') + \delta(\bm{r}-\bm{r}') V_C(r) , \nonumber \\
\nonumber
\end{eqnarray}

The nonlocal term is split into a volume and a narrower Gaussian term of opposite sign to make the final potential have a wine-bottle shape.

\begin{eqnarray}
\Sigma_{HF}^{nl}\left( \bm{r},\bm{r}' \right) = -V_{HF}^{vol}\left( \bm{r},\bm{r}'\right) 
+ V_{HF}^{wb}(\bm{r},\bm{r}') ,
\label{eq:HFn}
\nonumber
\end{eqnarray}
where the volume term is given by
\begin{equation}
\begin{split}
V_{HF}^{vol}\left( \bm{r},\bm{r}' \right) =  V^{HF}
\,f \left ( \tilde{r},r^{HF}_{(p,n)},a^{HF} \right ) \\ \times
 \left [ x H \left( \bm{s};\beta^{vol_1} \right) + (1-x) H \left( \bm{s};\beta^{vol_2}\right) \right ] \\
\end{split} \label{eq:HFvol} 
\end{equation}
allowing for two different nonlocalities with different weights ($0 \le x \le1$). 
With   the notation $\tilde{r} =(r+r')/2$ and $\bm{s}=\bm{r}-\bm{r}'$,
the wine-bottle ($wb$) shape is described by
\begin{equation}
V_{HF}^{wb}(\bm{r},\bm{r}') = V^{wb}_{(p,n)}  \exp{\left(- \tilde{r}^2/(\rho^{wb})^2\right)} H \left( \bm{s};\beta^{wb} \right ),
\label{eq:wb}
\end{equation}
where nonlocality is represented by a Gaussian form
\begin{equation}
H \left( \bm{s}; \beta \right) = \exp \left( - \bm{s}^2 / \beta^2 \right)/ (\pi^{3/2} \beta^3) .
\nonumber
\end{equation}
As usual, we employ a Woods-Saxon shape
\begin{eqnarray}
f(r,r_{i},a_{i})=\left[1+\exp \left({\frac{r-r_{i}A^{1/3}}{a_{i}}%
}\right)\right]^{-1} .
\label{Eq:WS}
\end{eqnarray}
The Coulomb term is obtained from the experimental charge density distribution for $^{40}$Ca~\cite{deVries:1987}.

The inclusion of additional high-energy proton reaction cross section data necessitated a more dynamic spin-orbit potential. To achieve this, we implemented a nonlocality contribution in the real potential:
\begin{equation}\begin{split}
      V^{nl}_{so}(\bm{r},\bm{r'})= \left( \frac{\hbar}{m_{\pi }c}\right)
^{2} V^{so}\frac{1}{\tilde{r}}\frac{d}{d\tilde{r}}f(\tilde{r},r^{so}_{(p,n)},a^{so})\; \bm{\ell}\cdot \bm{\sigma} \\
\times H(\bm{s};\beta^{so}), \end{split}
\label{eq:HFso}
\end{equation}
where $\left( \hbar /m_{\pi }c\right) ^{2}$=2.0~fm$^{2}$ 
as in Ref.~\cite{Mueller:2011}.

The introduction of nonlocality in the imaginary part of the self-energy is well-founded theoretically both for long-range correlations~\cite{Waldecker:2011} as well as in short-range ones~\cite{Dussan:2011}. 
Its implied $\ell$-dependence is essential in reproducing the correct particle number for protons and neutrons.
The assumed imaginary component of the potential has the form
\begin{eqnarray}
\textrm{Im}\ \Sigma( \bm{r},\bm{r}',E) =   \textrm{Im}\ \Sigma^{nl}(\bm{r},\bm{r}';E) + \delta(\bm{r}-\bm{r}') \mathcal{W}^{so}(r;E) .
\nonumber
\label{Eq:imsig}
\end{eqnarray}
The nonlocal contribution is represented by
\begin{eqnarray}
\label{eq:imnl}
&\textrm{Im}\ \Sigma^{nl}(\bm{r},\bm{r}';E) = \hspace{5cm} \\ \nonumber  
&-W^{vol}_{0\pm}(E) f\left(\tilde{r};r^{vol}_{\pm};a^{vol}_{\pm}\right)H \left( \bm{s}; \beta^{vol}\right) \hspace{1cm} \\ \nonumber
&+ 4a^{sur}W^{sur}_{\pm}\left( E\right)H \left( \bm{s}; \beta^{sur}\right) \frac{d}{d \tilde{r} }f(\tilde{r},r^{sur}_{\pm},a^{sur}_{\pm})  .
\end{eqnarray}
At energies well removed
from $\varepsilon_F$, the form of the imaginary volume potential should not be
symmetric about $\varepsilon_F$ as indicated by the $\pm$ notation in the subscripts and superscripts~\cite{Dussan:2011}.
While more symmetric about $\varepsilon_F$, we have allowed a similar option for the surface absorption that is also supported by theoretical work reported in Ref.~\cite{Waldecker:2011}.
We include a local spin-orbit contribution with the same form as in Eq.~(\ref{eq:HFso})
\begin{eqnarray}
\mathcal{W}^{so}(r,E)=  \left( \frac{\hbar}{m_{\pi }c}\right)
^{2}W^{so}(E) 
 \frac{1}{r}\frac{d}{dr}f(r,r^{so}_{(p,n)},a^{so})\; \bm{\ell}\cdot \bm{\sigma}, \nonumber \\
 \label{eq:IMso}
\end{eqnarray}
using the same geometry parameters as in Eq.~(\ref{eq:HFso}).
Allowing for the aforementioned asymmetry around $\varepsilon_F$ the following form was assumed for 
the depth of the volume potential~\cite{Mueller:2011}
\begin{widetext} 
\begin{equation}
W^{vol}_{0\pm}(E) =  \Delta W^{\pm}_{NM}(E) +  
\begin{cases}
0 & \text{if } |E-\varepsilon_F| < \mathcal{E}^{vol} \\
\left [ A^{vol} \pm \eta^{vol} \right ]  \frac{\left(|E-\varepsilon_F|-\mathcal{E}^{vol}\right)^4}
{\left(|E-\varepsilon_F|-\mathcal{E}^{vol}\right)^4 + (B^{vol})^4} & 
 \text{if } |E-\varepsilon_F| > \mathcal{E}^{vol} ,
\end{cases} 
\label{eq:volumeS}
\end{equation}
\end{widetext}
where $\Delta W^{\pm}_{NM}(E)$ is the energy-asymmetric correction modeled after
nuclear-matter calculations. The asymmetry above and below $\varepsilon_F$ is essential to accommodate the Jefferson Lab $(e,e'p)$ data at large missing energy.
The energy-asymmetric correction was taken as 
\begin{widetext} 
\begin{equation}
\Delta W^{\pm}_{NM}(E)=
\begin{cases}
\alpha \left [A^{vol}_{+} \pm \eta^{vol} \right ]\left[ \sqrt{E}+\frac{\left( \varepsilon_F+\mathbb{E}_{+}\right) ^{3/2}}{2E}-\frac{3}{2}
\sqrt{\varepsilon_F+\mathbb{E}_{+}}\right] & \text{for }E-\varepsilon_F>\mathbb{E}_{+} \\ 
- \left [ A^{vol}_{-} \pm \eta^{vol} \right ] \frac{(\varepsilon_F-E-\mathbb{E}_{-})^2}{(\varepsilon_F-E-\mathbb{E}_{-})^2+(\mathbb{E}_{-})^2} & \text{for }E-\varepsilon_{F}<-\mathbb{E}_{-} \\ 
0 & \text{otherwise}.
\end{cases} 
\label{eq:Wnmnl}
\end{equation} 
\end{widetext}

To describe the energy dependence of surface absorption we employed the form of Ref.~\cite{Charity:2007}.
\begin{eqnarray}
W^{sur}_{\pm}\left( E\right) =\omega _{4}(E,A^{sur},B^{sur_1},0)- \nonumber \\
\omega_{2}(E,A^{sur},B^{sur_2},C^{sur}),  \label{eq:paranl} 
\end{eqnarray}
where
\begin{eqnarray}
\omega _{n}(E,A^{sur},B^{sur},C^{sur})=A^{sur}\;\Theta \left(
X\right) \frac{X^{n}}{X^{n}+\left( B^{sur}\right) ^{n}}, \nonumber \\
\label{eq:omega}
\end{eqnarray}%
and $\Theta \left( X\right) $ is Heaviside's step function and $%
X=\left\vert E-\varepsilon_F\right\vert -C^{sur}$. 
As the imaginary spin-orbit component is
generally needed only at high energies, we have kept the form employed in Ref.~\cite{Mueller:2011}
\begin{equation}
W^{so}(E)= A^{so}  \frac{(E-\varepsilon_F)^4}{(E-\varepsilon_F)^4+(B^{so})^4} .
\label{eq:ImSO}
\end{equation}%
All ingredients of the self energy have now been identified and their functional form described.
In addition to the Hartree-Fock contribution and the absorptive potentials, we also include the dispersive real part from all imaginary contributions according to the corresponding subtracted dispersion relation (see Eq.~(\ref{eq:dispersion})).

\subsection*{Parameters}
\label{sec:Results}

The constraint of the number of particles was incorporated to include contributions from $\ell = 0$ to 5.
Such a range of $\ell$-values generates a sensible convergence with $\ell$ when short-range correlations are included as in Ref.~\cite{Dussan:2011}. 
We obtain 19.8 protons from all $\ell = 0$ to 5 partial wave terms including $j = \ell \pm \frac{1}{2}$ and 19.7 for neutrons.
This is within the error we assigned to the particle number of 1\%.
If in future higher $\ell$-values are included, we expect a slight but not essential change in the fitted parameters.   
The values of the fitted parameters are listed in Table~\ref{Tbl:fitted}.

\begin{table}[tpb]
\caption{Fitted parameter values for proton and neutron potentials in 
$^{40}$Ca. This table also lists the number of the equation that defines each individual parameter.}
\label{Tbl:fitted}%
\begin{ruledtabular}
\begin{tabular}{ccc}
parameter &                                  value & Eq. \\
\hline
\multicolumn{3}{c}{Hartree-Fock} \\
\hline
$V^{HF}$ [MeV]		& 93.6 & (\ref{eq:HFvol}) \\
$a^{HF}$ [fm]                       &  0.68   & (\ref{eq:HFvol}) \\
$\beta^{vol_1}$ [fm]       &  1.48     &(\ref{eq:HFvol}) \\
$\beta^{vol_2}$ [fm]       &  0.70     &(\ref{eq:HFvol}) \\
$x$ 				& 0.48	& (\ref{eq:HFvol}) \\
$\rho^{wb}$ [fm]                  &  0.69   & (\ref{eq:wb}) \\
$\beta^{wb}$ [fm]                 &  0.41   & (\ref{eq:wb}) \\
\hline
\multicolumn{3}{c}{Spin-orbit} \\
\hline
$V^{so} [MeV]$                    & 12.4    & (\ref{eq:HFso})\\
$a^{so}$ [fm]                       &  0.762   & (\ref{eq:HFso}) \\
$\beta^{so}$ [fm]       &  0.792     &(\ref{eq:HFso}) \\
$A^{so}$ [MeV]                     & -2.37 & (\ref{eq:ImSO}) \\
$B^{so}$ [MeV]                      &  31.8 &  (\ref{eq:ImSO})     \\
\hline
\multicolumn{3}{c}{Volume imaginary} \\
\hline
$a^{vol}_{+}$ [fm] 		& 0.698 	& (\ref{eq:imnl}) \\ 
$\beta^{vol}_{+}$ [fm]		& 1.15	& (\ref{eq:imnl}) \\ 
$a^{vol}_{-}$ [fm] 		& 0.470 	& (\ref{eq:imnl}) \\ 
$\beta^{vol}_{-}$	[fm]		& 0.26	& (\ref{eq:imnl}) \\
$A^{vol}_{+}$ [MeV]           & 6.61  & (\ref{eq:volumeS}) \\
$B^{vol}_{+}$ [MeV]		& 17.6	& (\ref{eq:volumeS}) \\
$\mathcal{E}^{vol}_{+}$ [MeV]	& 4.42	& (\ref{eq:volumeS}) \\
$A^{vol}_{-}$ [MeV]            & 17.4  & (\ref{eq:volumeS}) \\
$B^{vol}_{-}$ [MeV]		& 30.6	& (\ref{eq:volumeS}) \\
$\mathcal{E}^{vol}_{-}$ [MeV]		& 1.29	& (\ref{eq:volumeS}) \\
$\mathbb{E}_{+}$ [MeV]			& 25.0	& (\ref{eq:Wnmnl}) \\
$\mathbb{E}_{-}$ [MeV]			& 124	& (\ref{eq:Wnmnl}) \\
$\alpha$                      & 0.130 &(\ref{eq:Wnmnl}) \\
\hline
\multicolumn{3}{c}{Surface imaginary} \\
\hline
$a^{sur}_{+}$ [fm] 		& 0.688 	& (\ref{eq:imnl}) \\ 
$\beta^{sur}_{+}$ [fm]       & 3.38  & (\ref{eq:imnl}) \\ 
$a^{sur}_{-}$ [fm] 		& 1.48 	& (\ref{eq:imnl}) \\ 
$\beta^{sur}_{-}$ [fm]       & 1.72  & (\ref{eq:imnl}) \\ 
$A^{sur}_{+}$ [MeV]          & 14.2 & (\ref{eq:paranl}) \\
$B^{sur_1}_{+}$ [MeV]         & 5.21 & (\ref{eq:paranl}) \\
$B^{sur_2}_{+}$ [MeV]         & 145 & (\ref{eq:paranl}) \\
$C^{sur}_{+}$ [MeV] & 71.1   &  (\ref{eq:paranl}) \\
$A^{sur}_{-}$ [MeV]          & 5.33  & (\ref{eq:paranl}) \\
$B^{sur_1}_{-}$ [MeV]         & 9.73 & (\ref{eq:paranl}) \\
$B^{sur_2}_{-}$ [MeV]         & 30.2 & (\ref{eq:paranl}) \\
$C^{sur}_{-}$ [MeV] & 56.9 &  (\ref{eq:paranl}) \\
\end{tabular}
\end{ruledtabular}
\end{table}

\subsection*{Results}

We found the DOM self-energy by minimizing the $\chi^2$ using experimental data in the form of elastic-scattering cross sections, total and reaction cross sections, bound-state energy levels, charge density, and particle number. The results of this fit led to the curves shown in this supplementary material. The charge density and proton reaction cross section are reported in the main text.  

\begin{table}[tpb]
   \caption{Comparison of experimental and fitted mean energies for various proton and neutron orbitals for $^{40}$Ca.}
   \label{Tble:level}%
   \begin{ruledtabular}
      \begin{tabular}{ccccc}
         orbit      &  \multicolumn{2}{c}{Neutrons} & \multicolumn{2}{c}{Protons} \\
         &   Fitted   & Exp.  & Fitted & Exp. \\
         &  [MeV]  & [MeV] & [MeV] & [MeV] \\
         \hline
         0$d_{3/2}$ &  -15.4  & -15.6 & -8.13 & -8.3 \\
         0$d_{5/2}$ &  -21.72  & -22.3 & -14.4 & -14.3 \\
         1$s_{1/2}$ &  -16.5  & -18.3 &  -9.19 & -10.8 \\
         0$f_{7/2}$  &  -9.80 & -8.36   &  -2.85 & -1.09 \\
      \end{tabular}
   \end{ruledtabular}
\end{table}

\begin{figure}[h]
   \begin{minipage}{\columnwidth}
      \makebox[\columnwidth]{
         \includegraphics[scale=0.7]{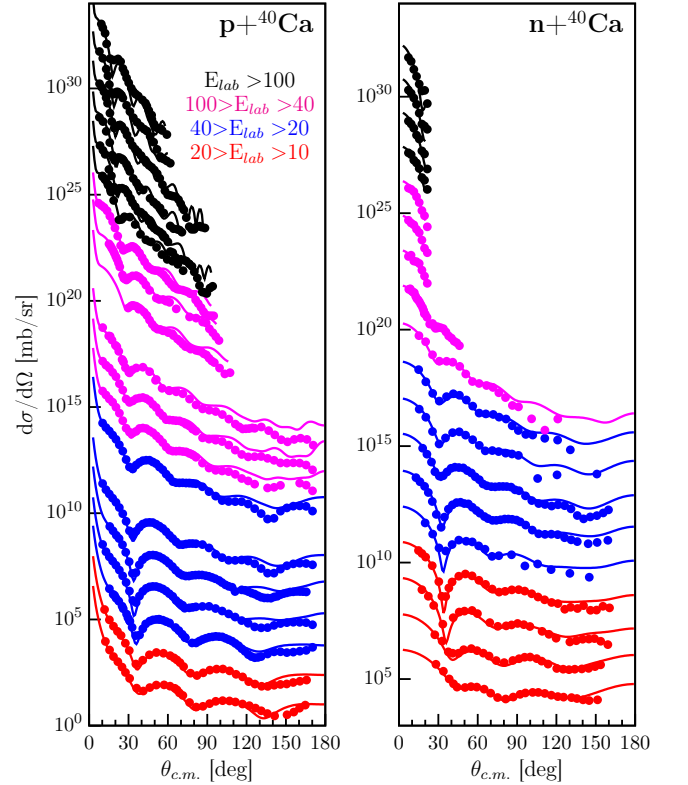}
      }
   \end{minipage}

   \caption{Calculated and experimental proton and neutron elastic-scattering angular distributions of the differential cross section $\frac{d\sigma}{d\Omega}$. The data at each energy is offset by factors of ten and plotted with a log scale to help visualize all of 
   the data at once. References to the data are given in~ \cite{Mueller:2011}.} 
   \label{fig:elastic}
\end{figure}

\begin{figure}[t]
   \begin{minipage}{\columnwidth}
      \makebox[\columnwidth]{
         \includegraphics[scale=0.7]{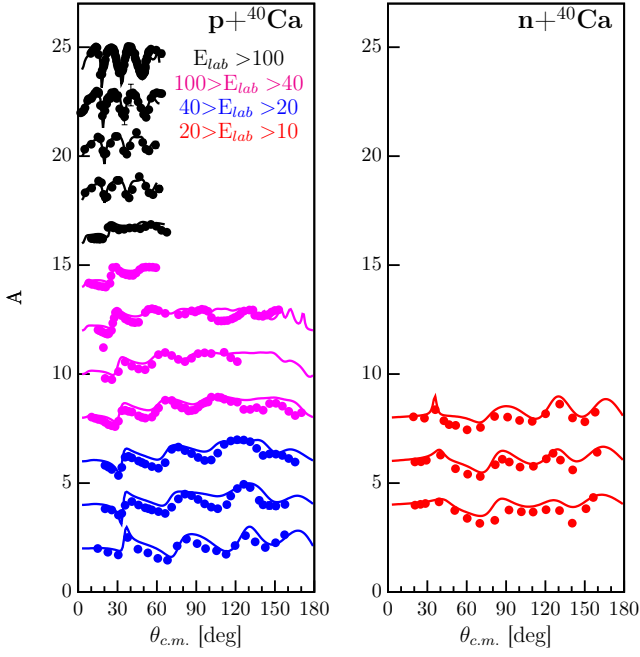}
      }
   \end{minipage}
   \caption{Results for proton and neutron analyzing power generated from the DOM self-energy. References to the data are given in~\cite{Mueller:2011}.}
   \label{fig:analyze}
\end{figure}

\begin{figure}[t]
   \begin{minipage}{\columnwidth}
      \makebox[\columnwidth]{
         \includegraphics[scale=0.6]{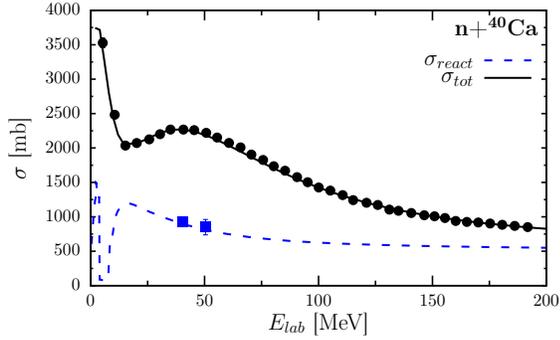}
      }
   \end{minipage}
   \caption{Neutron total cross section (solid line) and reaction cross section (dashed line) generated from the DOM self-energy. The circles represent measured total cross sections and the squares measured reaction cross sections. References to the data are given in~\cite{Mueller:2011}.}
   \label{fig:total}
\end{figure}

The resulting elastic-scattering cross sections are shown in Fig.~\ref{fig:elastic}, analyzing powers are shown in Fig.~\ref{fig:analyze}, and the neutron reaction and total cross sections results are shown in Fig.~\ref{fig:total}. Each fit is of the same quality as those of Refs.~\cite{Mueller:2011} and \cite{Charity:2007}.

Now that positive energies are accurately described by the DOM self-energy, the dispersion relation also constrains negative energy values. Negative energy information can be properly described due to the nonlocal implementation of the current analysis. This leads to the charge density detailed in Fig.~\ref{fig:chd}, particle numbers, as well as the bound-state energy levels, shown in Table \ref{Tble:level}.

\bibliographystyle{apsrev4-1}
\bibliography{eep}

\end{document}